\documentclass[twocolumn,
aps,
superscriptaddress,
pra
]{revtex4-1}

\usepackage[utf8]{inputenc}
\usepackage[T1]{fontenc}
\usepackage{graphicx}
\usepackage{wrapfig}
\usepackage{booktabs}
\usepackage{amsmath}
\usepackage{amssymb}
\usepackage{braket}
\usepackage[dvipsnames]{xcolor}
\usepackage[normalem]{ulem}
\usepackage{bbold}

\usepackage[colorlinks=true,urlcolor=blue,citecolor=blue,linkcolor=red]{hyperref}

\begin{document}
	
	\title{Insensitivity points and performance of open quantum interferometers \\ under number- conserving \& non-conserving Lindblad dynamics}	
	
	\author{Tommaso Favalli}\email{tommaso.favalli@units.it}
	\affiliation{University of Trieste, Strada Costiera 11, I-34151 Trieste, Italy}
	\author{\v{Z}an Kokalj}\email{zan.kokalj@phd.units.it}
	\affiliation{University of Trieste, Strada Costiera 11, I-34151 Trieste, Italy}
	\author{Andrea Trombettoni}\email{atrombettoni@units.it}
	\affiliation{University of Trieste, Strada Costiera 11, I-34151 Trieste, Italy}
	
	\begin{abstract}
		We investigate the phase sensitivity of a linear two-mode atom interferometer subject to environmental noise, modeled within the framework of open quantum systems with both number- conserving and non-conserving Lindblad operators. Considering several input states, we first study the cases $N=1,2$ ($N$ number of particles) and perform numerical simulations for $N>2$. The sensitivity as a function of the holding time can display divergence points where phase estimation becomes impossible, to which we refer to as insensitivity points. We characterize their behavior as the input state, particle number, and noise operator are varied, and we find that their positions are independent of the noise intensity. Moreover, while our fixed measurement scheme may favor number-conserving noise at small $N$ (i.e., having better sensitivity), the Cram\'{e}r–Rao bound reveals that particle non-conserving noise yields strictly lower achievable sensitivity for all particle numbers. 
	\end{abstract}
	
	\maketitle
	
	\section{Introduction}
	Since realistic sensing devices are unavoidably exposed to noise, decoherence, and different sources of loss, a large body of research has investigated how these effects limit their sensitivity \cite{degen2017quantum}. This activity has been applied in particular to quantum interferometers, due both to their paradigmatic role as quantum sensors \cite{pezze2018quantum} and to their broad range of applications \cite{Cronin2009,abend2020atom,jin2024quantum}. A considerable amount of work has been devoted to the topic of quantifying and characterizing the effect of different kinds of noise on quantum interferometers \cite{demkowiczdobrzanski2009quantum,Davidovich2011,escher2012quantum,das2012universal,DD2012,demkowiczdobrzanski2014using,yue2014quantum,kolthesis,demkowiczdobrzanski2017adaptive,albarelli2018restoring,albarelli2022quantum,gorecki2022quantum,bao2022multichannel,gorecki2025interplay,das2025quantum,kurdzialek2025universal}. These studies address the behavior of the sensitivity, develop tools to characterize the impact of noise, and propose strategies to mitigate decoherence and improve performance as close as possible to the fundamental precision limits.
	
	Such a characterization is particularly relevant both for emerging quantum technologies -- where identifying operational regimes that avoid severe degradation of sensitivity is essential -- and for fundamental investigations \cite{degen2017quantum,pezze2018quantum,Alonso2022}. In recent years, quantum sensors based on atom interferometry have begun to reach a level of reliability and robustness suitable for deployment beyond laboratory environments. Instruments tailored for geophysical and space-based measurements \cite{Freier_2016,Menoret_2018,cooke2021,kanxing_2021,bidel2023airborne,Chen_2023,antonimicollier2024absolute} are now being used for gravimetry and inertial navigation \cite{ExailHomepage}. Related platforms have enabled precision measurements of magnetic fields \cite{Rovny2022NanoscaleCovarianceMagnetometry} and the gravity vector in optical lattices \cite{LeDesma2025}, illustrating the rapid expansion of quantum metrology into practical and fundamental domains.
	
	A general framework for establishing precision bounds in noisy interferometric settings has been developed in \cite{Davidovich2011}, and considerable effort has been devoted to designing strategies capable of counteracting the detrimental impact of noise and loss \cite{demkowiczdobrzanski2014using,albarelli2018restoring,albarelli2022quantum}. The microscopic description of specific noise or loss mechanisms has been investigated thoroughly \cite{demkowiczdobrzanski2009quantum,demkowiczdobrzanski2017adaptive,yue2014quantum,gorecki2022quantum,kurdzialek2025universal}. A central insight emerging is that, in the presence of Markovian noise, the attainable sensitivity often departs from the ideal Heisenberg scaling and becomes constrained by the structure of the dissipative dynamics. Optimal protocols in noisy metrology generally require careful tailoring of probe states together with adaptive control operations \cite{demkowiczdobrzanski2014using}, and error-corrected metrological schemes can be constructed systematically by identifying subspaces in which the parameter-imprinting Hamiltonian is incompatible with the noise operators \cite{albarelli2018restoring}. There is in general an interplay between the parameter encoding and the Lindbladian structure of the environment that determines whether quantum enhancements can be preserved or partially recovered in realistic settings.
	
	To reach this goal, a detailed understanding of how concrete physical noise processes affect quantum interferometry has to be developed. A possible approach is to model specific decoherence channels—such as dephasing, phase diffusion, photon or particle loss—directly at the microscopic level, identifying how each mechanism degrades the attainable precision \cite{demkowiczdobrzanski2009quantum}. By introducing adaptive strategies and realistic feedback protocols, the interplay between coherent dynamics and dissipation was analysed to constrain metrological performance under optimized control schemes \cite{demkowiczdobrzanski2017adaptive}. Noise sources arising from technical or experimental imperfections, including fluctuating phase imbalances or path instabilities, were also investigated, and it was shown how such mechanisms can be mapped onto effective stochastic processes directly influencing the quantum Fisher information \cite{yue2014quantum}.
	
	Altogether, a significant amount of literature aims at providing a comprehensive microscopic characterisation of noise processes that can arise in practical implementations of quantum interferometers. Here we follow the approach of modelling the noise through a Lindblad dynamics, and we analyse the interferometer through the behavior of its sensitivity. As is well known for a linear two-mode Mach–Zehnder interferometer, reconstructing the parameter to be estimated becomes more difficult—or even impossible—at points where the expectation value of the measured observable at the output depends only weakly (or not at all) on the parameter itself.
	
	A simple way to see this is to consider that, at the end of the interferometric process, one measures an observable $\hat{O}$ and extracts the parameter to be estimated, say $\delta$, with a sensitivity \cite{pezze2018quantum}:
	\begin{equation}
		\Delta \delta = \frac{\Delta \hat{O}}{\left| \partial \langle \hat{O} \rangle / \partial \delta \right| } .
	\end{equation}
	When the denominator vanishes, the sensitivity is completely degraded. For example, this may occur—fixing all other parameters—at specific holding times between the two beam-splitter operations performed during the interferometric sequence.
	
	We refer to these points as \emph{insensitivity points}: the times at which phase estimation becomes impossible. Of course, one should avoid such points as much as possible, since they place the system far from the precision bounds. Rather than treating them as pathologies of the open-system dynamics, we use their structure, abundance, and robustness as operational indicators of when an interferometric scheme loses metrological usefulness. This perspective provides guidance for assessing and designing realistic atom–interferometric protocols, by identifying parameter regions and time intervals in which the system should avoid operating in order to retain better sensitivity. In particular, we study how insensitivity points depend on the strength of the environmental coupling and on the nature of the environment itself, by investigating two types of Lindblad operators: number-conserving and non-conserving.
	
	We focus on these two classes since they are associated with different and representative physical mechanisms (dephasing and dissipation) of noise emerging from the coupling with an environment \cite{libroBP}. Indeed, we may expect number–non–conserving channels to degrade the sensitivity more severely, since they act more invasively on states with quantum correlations such as cat states \cite{pezze2018quantum}. However, our results show a different behavior. For the measurement scheme adopted here, the sensitivity obtained with number-conserving noise is better at small particle numbers, whereas particle non-conserving noise becomes advantageous for larger $N$. When instead considering the Cram\'{e}r–Rao lower bound (CRLB), the non-conserving case performs better in terms of sensitivity with respect to the number-conserving one for all $N$.
	
	The paper is organized as follows. In Sec.~II we introduce the model used to describe our open quantum interferometer, while in Sec.~III we outline the interferometric protocol and the role of the different dynamical stages. Secs.~IV and~V are devoted to the analytical treatment of the cases with one and two particles, respectively. In Sec.~VI we extend the analysis to the regime $N>2$ ($N$ number of particles), relying on numerical simulations to investigate how the behavior observed in the few-particle limit generalizes to larger systems. Finally, in Sec.~VII we present our conclusions.

	\section{The model}	
	\label{sec:II}
	As a physical system of reference, we deal with a condensate of bosons in a double-well potential \cite{smerzi1997quantum,der2,der1,der3,BH}, with the possibility to tune the height of the barrier and the energy difference between the wells \cite{morsch2006dynamics}. One can use an optical lattice with many energy barriers 
	\cite{jaksch1998cold,trombettoni2001discrete,cataliotti2001josephson,morsch2006dynamics}. The simple setup of (possibly interacting) ultracold bosons gives rise to an interferometer, where the splitting process is realized by letting the barrier become high enough for the atoms in the two wells to be isolated, in order to accumulate a phase difference. After this accumulation stage, recombination is performed by lowering the barrier. A useful tool to describe bosons in a double-well potential is the two-mode approximation \cite{smerzi1997quantum,der1}, consisting in neglecting all energy levels but the first two and rearranging them in such a way that the system can be described by two effective modes corresponding to the two wells $a$ and $b$. In this way, the full many-body description of an interacting condensate in a double-well potential is simplified to a two-site Bose–Hubbard model with the Hamiltonian \cite{der1}:
	\begin{equation}\label{BH}
		\hat{H} = -J(t)\left(\hat{a}^{\dagger} \hat{b} + \hat{b}^{\dagger} \hat{a}\right)
		+ \frac{\delta(t)}{2}\left(\hat{a}^{\dagger} \hat{a} - \hat{b}^{\dagger} \hat{b}\right) ,
	\end{equation}
	where $\hat{a}^{\dagger},\hat{a}$ ($\hat{b}^{\dagger},\hat{b}$) are the creation/annihilation operators for the left (right) well, $J(t)$ is the tunneling strength, $\delta(t)$ is the energy shift between the two wells, and where the interaction term $\frac{U}{2}\left(\hat{a}^{\dagger}\hat{a}^{\dagger}\hat{a}\hat{a} + \hat{b}^{\dagger}\hat{b}^{\dagger}\hat{b}\hat{b}\right)$
	has been omitted since we do not consider interactions within the wells in this work. The functions $J(t)$ and $\delta(t)$ are written as time-dependent in the sense that they can be switched on and off during the various stages of the interferometric process. We will denote the basis states of the Fock space by $\{\ket{n, N-n}\}$, with $n = 0, 1, \cdots, N$, where $\ket{n, N-n}$ has $n$ particles in mode $a$ and $N-n$ in mode $b$.
	
	In our discussion we will consider the interferometer as an open quantum system interacting with an environment. In doing this, we will assume that such noise is present only during the phase-accumulation stage, when the tunneling term $J(t)$ is zero. In Appendix A we study the interferometer dynamics with noise present throughout the whole process, comparing the results with those presented in the main text and showing that they are only slightly modified quantitatively and not qualitatively.
	
	Under a well-known set of suitable conditions and approximations (that we do not discuss here; see, for example, the discussion in \cite{libroBP}), the non-unitary evolution of the particles in our open quantum interferometer will be described by the Lindblad master equation ($\hslash=1$):
	\begin{equation}\label{LINDBLAD}
		\frac{\partial \rho}{\partial t} = - i\left[\hat{H},\rho\right] + \mathcal{L}[\rho] ,
	\end{equation}
	where
	\begin{equation}
		\mathcal{L}[\rho] =
		\sum_{j} \gamma_j \left(
		\hat{L}_j \rho \hat{L}^{\dagger}_j
		- \frac{1}{2}\left\{\hat{L}^{\dagger}_j \hat{L}_j , \rho\right\}
		\right)
	\end{equation}
	with $\hat{L}_j$ operators describing the dissipative part of the dynamics and $\gamma_j$ a set of non-negative coefficients providing the strength of the coupling \cite{lindblad1,lindblad2}. 
	
	We will focus on three input states: $N0 \equiv \ket{N,0}$, $TF \equiv \ket{N/2,N/2}$, and $NOON \equiv \frac{1}{\sqrt{2}}(\ket{N,0}+\ket{0,N})$. In the noiseless case, the first yields shot-noise sensitivity, while the latter two achieve Heisenberg-limited scaling \cite{Pezz__2009,pezze2018quantum}. In analysing the effect of noise, we will consider three number-conserving Lindblad operators, namely $\hat{S}_{z}$ and $\hat{S}_{\pm}$, while in Sec.~VI we also examine—and explicitly compare with these—a particle non-conserving example.
	
	Through analytical calculations (for $N=1,2$) and numerical simulations (for $N>2$), we study in the next sections the behavior of the insensitivity points and show that, while such divergences already occur in some noiseless cases, the presence of noise systematically generates additional ones. These points appear periodically and, remarkably, their positions are entirely unaffected by the noise strength: changing $\gamma_j$ modifies the overall sensitivity, but never shifts the times at which divergences occur. This robustness is one of the central findings of our work. Moreover, we show that the number of insensitivity points either grows linearly with the number of particles $N$ or remains constant, depending on the input state, thus providing an additional criterion for assessing the experimental feasibility of different interferometric configurations. Finally, we investigate how the optimal phase sensitivity—i.e.\ that obtained away from the insensitivity points—scales with the particle number in the presence of noise, comparing it with the corresponding noiseless behavior.

	\section{The interferometric process}
	In this section we describe the complete interferometric sequence, consisting in a first beam splitter, a phase accumulation stage and a second beam splitter. As previously mentioned, in the main text we will assume that the noise acts only during the phase accumulation stage. 
	
	The dynamics of our open quantum interferometer is thus characterized by the following steps:
	\begin{itemize}
		\item{$0<t<T_{BS}$:} The system undergoes the unitary evolution
		\begin{equation}\label{evut}
			\frac{\partial \rho}{\partial t} = - i\left[\hat{H}_J,\rho\right]
		\end{equation}
		where the Hamiltonian is given by:
		\begin{equation}\label{HJ}
			\hat{H}_J = - J \left(\hat{a}^{\dagger} \hat{b} + \hat{b}^{\dagger} \hat{a}\right) .
		\end{equation} 
		In this stage the particles are free to tunnel through the barrier in a symmetric double well.
		
		\item{$T_{BS}<t<T_{BS} + T_H$:} The particle tunneling is switched off and an asymmetry is introduced among the wells by setting $\delta \ne 0$. The system is here governed by the Lindblad master equation
		\begin{equation}\label{evlind}
			\frac{\partial \rho}{\partial t} = - i\left[\hat{H}_{\delta},\rho\right] + \gamma \hat{L} \rho \hat{L}^{\dagger} - \frac{\gamma}{2}\left\{\hat{L}^{\dagger} \hat{L} ,\rho\right\}
		\end{equation}
		where the Hamiltonian is now:
		\begin{equation}\label{Hdelta}
			\hat{H}_{\delta} =  \frac{\delta}{2}\left(\hat{a}^{\dagger} \hat{a} - \hat{b}^{\dagger} \hat{b}\right).
		\end{equation}
		In this stage the system evolves for an arbitrary holding time $T_H$, in which a phase is accumulated due to the asymmetry between the wells.
		
		\item{$T_{BS} + T_H<t<T_H + 2T_{BS}$:} The symmetry is restored by taking again $\delta=0$ and the second beam splitter is implemented allowing the particle to tunnel for a time $T_{BS}$. The evolution of the system is governed again by (\ref{evut}) with the Hamiltonian (\ref{HJ}).
	\end{itemize}
	
	Assuming the time is scaled by $1/J$ (with $J=1$ for simplicity), we take $T_{BS}=\pi/4$. At the end of the interferometric process the sensitivity will be calculated as a function of the holding time $T_H$ and we will show how it behaves depending on $\gamma$, the input state and $N$.

	\section{Open interferometer with $N=1$}
	In this section we study the phase-sensitivity when only one particle is present within the interferometer. The three noise operators $\hat{S}_{z}$ and $\hat{S}_{\pm}$ lead to the same results. Thus, in order not to complicate the discussion, we show calculations simultaneously.

	\subsection{Input state $N0=\ket{1,0}$}
	For the input state $N0=\ket{1,0}$ we have the following initial density matrix:
	\begin{equation}\label{rho0}
		\rho_0 = \left(\begin{matrix}
			1&0\\0&0
		\end{matrix}\right)
	\end{equation}
	which, in the absence of noise, leads to sensitivity $\Delta \delta = 1/T_H$. This is our reference behavior for the sensitivity that we will compare with the results we get assuming the open quantum interferometer, where the noise is present.
	
	In the first part of the interferometric process the system undergoes the unitary evolution (\ref{evut}) governed by the Hamiltonian ($J=1$):
	\begin{equation}
		\hat{H}_J = \left( \begin{matrix}
			0 & -1 \\ -1 & 0
		\end{matrix}\right) .
	\end{equation}
	Solving the differential equation $\frac{\partial \rho}{\partial t} = - i\left[\hat{H}_J,\rho\right]$, imposing (\ref{rho0}) as initial condition and taking $t=T_{BS}=\pi/4$, we find:
	\begin{equation}\label{rhoTB}
		\rho(T_{BS}) = \left(\begin{matrix}
			1/2 & -i/2 \\ i/2 & 1/2
		\end{matrix}\right)
	\end{equation}
	which represents the initial state for the phase accumulation stage where the tunneling strength is set to zero and the asymmetry between the wells is introduced.
	
	We start considering $\hat{S}_z$ as noise operator, defined: $\hat{S}_z = \frac{1}{2}\left( \hat{a}^{\dagger}\hat{a} - \hat{b}^{\dagger}\hat{b}\right)$. In the basis of occupation numbers, such operator is represented by the matrix:
	\begin{equation}
		\hat{S}_z = \frac{1}{2} \left(\begin{matrix}
			1&0\\0&-1
		\end{matrix}\right) . 
	\end{equation} 
	Equation (\ref{evlind}), governing the dynamics of the system in this stage, becomes:
	\begin{equation}\label{evSz}
			\frac{\partial \rho}{\partial t} 
			= - i\left[\hat{H}_{\delta},\rho\right] + \gamma \hat{S}_z \rho \hat{S}_z - \frac{\gamma}{4} \rho 
	\end{equation}
	where the Hamiltonian is given by: 
	\begin{equation}\label{13}
		\hat{H}_{\delta} =  \left(\begin{matrix}
			\delta/2&0\\0&-\delta/2
		\end{matrix}\right) . 
	\end{equation} 
	Using now (\ref{rhoTB}) as initial state in solving the differential equation, we find (after the holding time $T_H$):
	\begin{equation}\label{rhoTBTH}
		\rho^{(z)}(T_{BS}+ T_H) = \left(\begin{matrix}
			1/2 & \frac{-i}{2} e^{-i\delta T_H}e^{- \frac{\gamma}{2}T_H} \\ \frac{i}{2} e^{i\delta T_H}e^{- \frac{\gamma}{2} T_H} & 1/2
		\end{matrix}\right) .
	\end{equation}
	This density matrix describes the state of the particle in the interferometer with $\hat{L}=\hat{S}_z$ before the last step.
	
	Considering instead $\hat{L}=\hat{S}_{-}$, the evolution of the particle in the asymmetric double well is given by:
	\begin{equation}
		\frac{\partial \rho}{\partial t} =  - i\left[\hat{H}_{\delta},\rho\right] + \gamma \hat{S}_{-} \rho\hat{S}_{+} - \frac{\gamma}{2}\hat{S}_{+}\hat{S}_{-}\rho -\frac{\gamma}{2} \rho\hat{S}_{+}\hat{S}_{-}
	\end{equation} 
	with the Hamiltonian given again by (\ref{13}). Solving this equation with (\ref{rhoTB}) as initial state and letting the system evolve for a time $T_H$, we obtain:
	\begin{equation}\label{r-}
		\rho^{(-)}(T_{BS} + T_H) = \left(\begin{matrix}
			\frac{1}{2}   e^{-\gamma T_H}& \frac{-i}{2} e^{-i\delta T_H}e^{- \frac{\gamma}{2}T_H} \\ \frac{i}{2} e^{i\delta T_H}e^{- \frac{\gamma}{2} T_H} &	\frac{1}{2}  \left(2- e^{-\gamma T_H}\right) 
		\end{matrix}\right) .
	\end{equation}
	Similarly, for $\hat{L}=\hat{S}_{+}$, the evolution during the phase accumulation stage is
	\begin{equation}
		\frac{\partial \rho}{\partial t} =  - i\left[\hat{H}_{\delta},\rho\right] + \gamma \hat{S}_{+} \rho\hat{S}_{-} - \frac{\gamma}{2}\hat{S}_{-}\hat{S}_{+}\rho -\frac{\gamma}{2} \rho\hat{S}_{-}\hat{S}_{+}
	\end{equation}
	leading to:
	\begin{equation}\label{r+}
		\rho^{(+)}(T_{BS} + T_H) = \left(\begin{matrix}
			\frac{1}{2}  \left(2- e^{-\gamma T_H}\right) & \frac{-i}{2} e^{-i\delta T_H}e^{- \frac{\gamma}{2}T_H} \\ \frac{i}{2} e^{i\delta T_H}e^{- \frac{\gamma}{2} T_H} & \frac{1}{2}   e^{-\gamma T_H}
		\end{matrix}\right) .
	\end{equation}
	We notice that the only difference between the density matrices (\ref{r-}) and (\ref{r+}) is a swithing in the diagonal terms, while the off-diagonal terms are exactly the same. We also notice that the off-diagonal terms are the equal to those encountered in the case of $\hat{L}=\hat{S}_z$.
	
	During the last part of the process the particle is again free to tunnel through the barrier and the unitary dynamics of the system is again governed by the Hamiltonian (\ref{HJ}). Assuming thus (\ref{rhoTBTH}), (\ref{r-}) and (\ref{r+}) as initial states, solving the differential equations and substituting $T_{BS}=\pi/4$, we obtain as final states:
	\begin{equation}
		\rho^{(z)}_f = \left(\begin{matrix}
			\frac{1}{2} - \frac{1}{2}  e^{-\frac{\gamma}{2}T_H} \cos(T_H\delta) & - \frac{1}{2}  e^{-\frac{\gamma}{2}T_H} \sin(T_H\delta) \\ - \frac{1}{2}  e^{-\frac{\gamma}{2}T_H} \sin(T_H\delta) & \frac{1}{2} + \frac{1}{2}  e^{-\frac{\gamma}{2}T_H} \cos(T_H\delta)
		\end{matrix}\right)
	\end{equation}
	and
	\begin{widetext}
		\begin{equation}
			\rho^{(\pm)}_f = \left(\begin{matrix}
				\frac{1}{2} - \frac{1}{2}  e^{-\frac{\gamma}{2}T_H} \cos(T_H\delta) & - \frac{1}{2}  e^{-\frac{\gamma}{2}T_H} \left( \sin(T_H\delta) \pm 2i \sinh\left(\frac{\gamma T_H}{2}\right) \right) \\ - \frac{1}{2}  e^{-\frac{\gamma}{2}T_H} \left( \sin(T_H\delta) \mp 2i \sinh\left(\frac{\gamma T_H}{2}\right) \right) & \frac{1}{2} + \frac{1}{2}  e^{-\frac{\gamma}{2}T_H} \cos(T_H\delta)
			\end{matrix}\right) .
		\end{equation}
	\end{widetext}
	
\begin{figure}[t!] 
		\centering 
		\includegraphics [height=13cm]{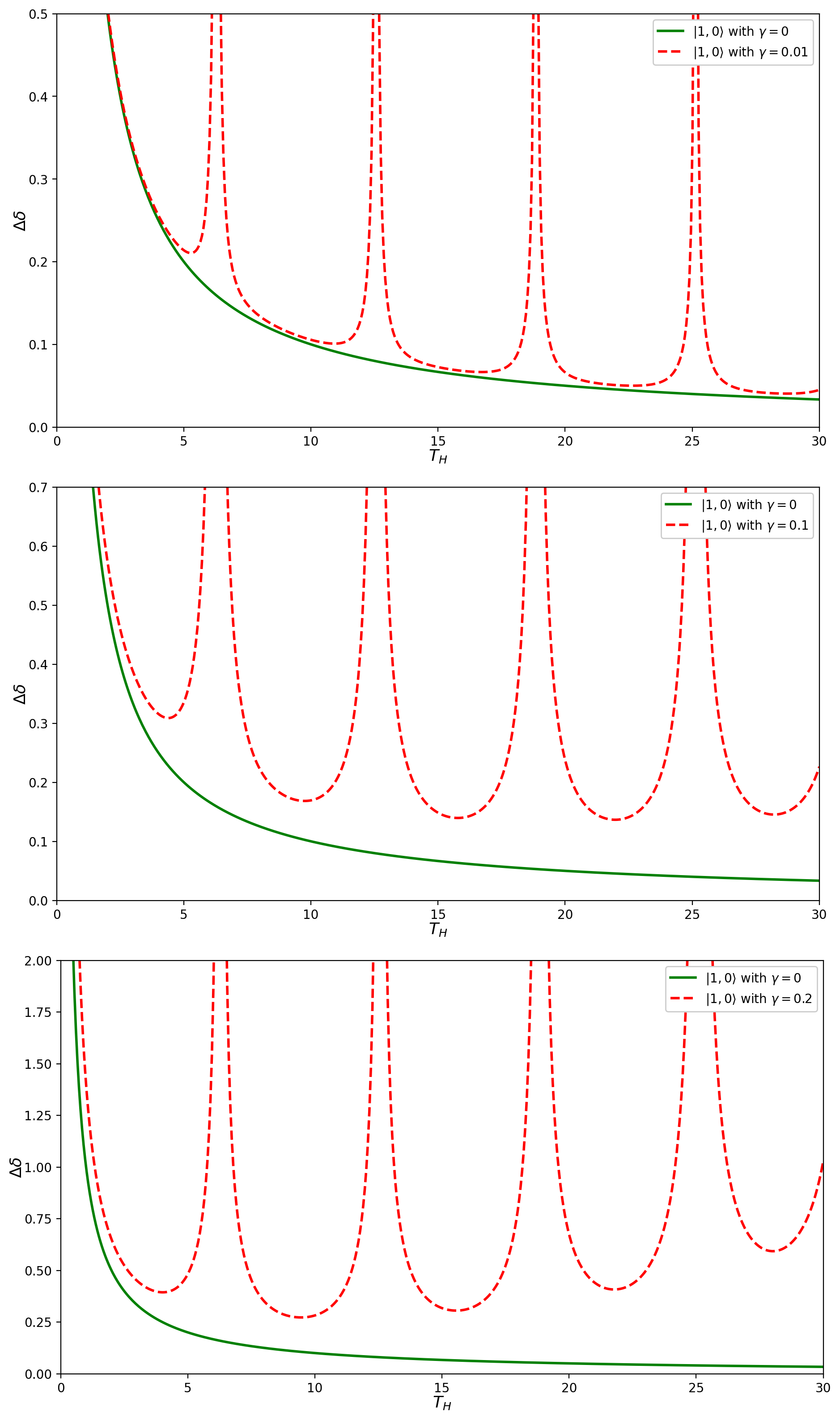} 
		\caption{The sensitivity of the interferometer for input state $\ket{1,0}$, noise operators $\hat{S}_{z}$, $\hat{S}_{\pm}$ and $N=1$ is plotted as a function of $T_H$ for $\delta=0.5$ and different values of $\gamma$. The solid lines show the noiseless case function $1/T_H$.} 
		\label{FIG1} 
	\end{figure}

	In order to find the sensitivity we choose here to calculate the expectation value, the variance and the $\delta$-derivative of the operator $\hat{n}=\hat{n}_b - \hat{n}_a = \hat{b}^{\dagger}\hat{b} - \hat{a}^{\dagger}\hat{a}$.
	In this way we have: $\Delta \delta = \frac{\Delta \hat{n}}{\left| \partial \langle \hat{n} \rangle / \partial \delta\right|}$. For all the noise operators, 
	namely starting from $\rho^{(z)}_f$ and $\rho^{(\pm)}_f$, we obtain:
	\begin{equation}\label{oioi}
		\left\{
		\begin{array}{l}
			\langle \hat{n} \rangle
			= e^{-\frac{\gamma}{2}T_H}\cos(T_H\delta)
			\\[6pt]
			
			\frac{\partial \langle \hat{n} \rangle}{\partial \delta}
			= - T_H e^{-\frac{\gamma}{2}T_H}\sin(T_H\delta)
			\\[6pt]
			
			\Delta \hat{n}
			= \sqrt{1 - e^{- \gamma T_H}\cos^{2}(T_H\delta)}
		\end{array}
		\right.
	\end{equation}
	where we used
	\begin{equation}
		\Delta \hat{n}
		= \sqrt{ \langle \hat{n}^2 \rangle - \langle \hat{n} \rangle^2 } \, .
	\end{equation}
	From (\ref{oioi}) we derive:
	\begin{equation}\label{sensN=1}
		\Delta \delta = \frac{\sqrt{1- e^{- \gamma T_H}\cos^{2}(T_H\delta)}}{\left| - T_H e^{-\frac{\gamma}{2}T_H}\sin(T_H\delta) \right|} .
	\end{equation}
	In Fig.~\ref{FIG1} the sensitivity is plotted as a function of the holding time $T_H$ for different values of $\gamma$. As could be expected, the noise increasingly degrades the sensitivity as $\gamma$ increases and, for $\gamma \rightarrow 0$, the sensitivity tends to the noise-free function $1/T_H$. We can notice the presence of what we call insensitivity points, namely the values of time for which $\Delta\delta$ diverges because of the sine function in the denominator. Such points occur for $T_H=\pi n/\delta$ ($n \in \mathbb{N}$) and do not depend on the value of $\gamma$. 
	
	\subsection{Input state $NOON$}
	For the input state $NOON=\frac{1}{\sqrt{2}}\left(\ket{1,0}+\ket{0,1}\right)$ we have the following initial density matrix:
	\begin{equation}\label{rho0NOON}
		\rho_0 = \frac{1}{2}\left(\begin{matrix}
			1&1\\1&1
		\end{matrix}\right)
	\end{equation}
	which, in the absence of noise, leads again to $\Delta \delta = 1/T_H$.
	
	In the first part of the interferometric process the system undergoes the unitary evolution (\ref{evut}) governed by the Hamiltonian $\hat{H}_J$. Solving the differential equation $\frac{\partial \rho}{\partial t} = - i\left[\hat{H}_J,\rho\right]$ and imposing (\ref{rho0NOON}) as initial condition, we find (after $t=T_{BS}=\pi/4$): $\rho(T_{BS}) \equiv \rho_0$. 
	The state is thus not changed by the first beam splitter.
	The dynamics of the interferometer during the phase accumulation stage is again described by equation (\ref{evlind}). After the holding time $T_H$ has passed we find:
	\begin{equation}
		\rho^{(z)}(T_{BS}+ T_H) = \left(\begin{matrix} 1/2 & \frac{1}{2} e^{-i\delta T_H}e^{- \frac{\gamma}{2}T_H} \\ \frac{1}{2} e^{i\delta T_H}e^{- \frac{\gamma}{2} T_H} & 1/2\end{matrix}\right) \label{rhoTBTHNOON} .
	\end{equation}
	and
	\begin{align}
		\rho^{(-)}(T_{BS}+ T_H) &= \left(\begin{matrix}
			\frac{1}{2} e^{-\gamma T_H} & \frac{1}{2} e^{-i\delta T_H}e^{- \frac{\gamma}{2}T_H} \\ \frac{1}{2} e^{i\delta T_H}e^{- \frac{\gamma}{2} T_H} & \frac{1}{2} \left(2 - e^{-\gamma T_H} \right)
		\end{matrix}\right) \label{rhoTBTHNOON-} \\
		\rho^{(+)}(T_{BS}+ T_H) &= \left(\begin{matrix}
			\frac{1}{2} \left(2 - e^{-\gamma T_H}\right) & \frac{1}{2} e^{-i\delta T_H}e^{- \frac{\gamma}{2}T_H} \\ \frac{1}{2} e^{i\delta T_H}e^{- \frac{\gamma}{2} T_H} & \frac{1}{2} e^{-\gamma T_H}
		\end{matrix}\right) \label{rhoTBTHNOON+}.
	\end{align}
	These density matrices describe the states of the particle in the interferometer before the second beam splitter.
	
\begin{figure}[t!] 
		\centering 
		\includegraphics [height=13cm]{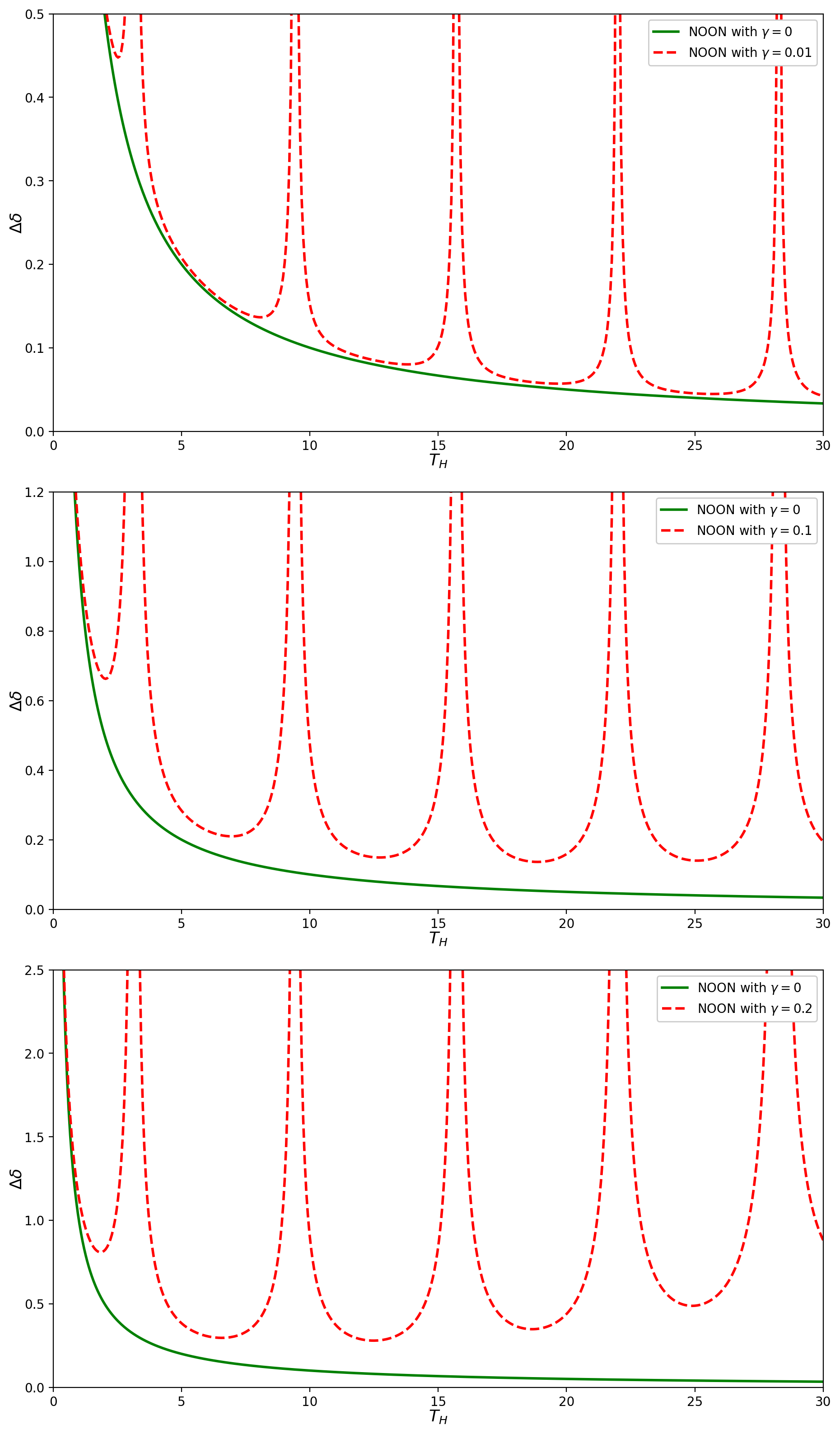} 
		\caption{The sensitivity of the interferometer for input state $NOON$, noise operators $\hat{S}_{z}$, $\hat{S}_{\pm}$ and $N=1$ is plotted as a function of $T_H$ for $\delta=0.5$ and different values of $\gamma$. The solid lines show the noiseless case function $1/T_H$.} 
		\label{FIG2} 
	\end{figure}

	\begin{widetext}
		During the last part of the process the particle is again free to tunnel through the barrier and the unitary dynamics of the system is governed by the Hamiltonian (\ref{HJ}). Assuming thus (\ref{rhoTBTHNOON}), (\ref{rhoTBTHNOON-}) and (\ref{rhoTBTHNOON+}) as initial states, solving the differential equations and substituting $T_{BS}=\pi/4$, we obtain as final states:
		\begin{equation}\label{rhofNOON}
			\rho^{(z)}_f = \left(\begin{matrix}
				\frac{1}{2} - \frac{1}{2}  e^{-\frac{\gamma}{2}T_H} \sin(T_H\delta) & \frac{1}{2}  e^{-\frac{\gamma}{2}T_H} \cos(T_H\delta) \\  \frac{1}{2}  e^{-\frac{\gamma}{2}T_H} \cos(T_H\delta) & \frac{1}{2} + \frac{1}{2}  e^{-\frac{\gamma}{2}T_H} \sin(T_H\delta)
			\end{matrix}\right) 
		\end{equation}
		and
		\begin{equation}\label{rhofNOONpm}
			\rho^{(\pm)}_f = \left(\begin{matrix}
				\frac{1}{2} - \frac{1}{2}  e^{-\frac{\gamma}{2}T_H} \sin(T_H\delta) & \frac{1}{2}  e^{-\frac{\gamma}{2}T_H} \left( \cos(T_H\delta) \mp 2i \sinh\left(\frac{\gamma T_H}{2}\right) \right) \\  \frac{1}{2}  e^{-\frac{\gamma}{2}T_H} \left( \cos(T_H\delta) \pm 2i \sinh\left(\frac{\gamma T_H}{2}\right) \right) & \frac{1}{2} + \frac{1}{2}  e^{-\frac{\gamma}{2}T_H} \sin(T_H\delta)
			\end{matrix}\right) .
		\end{equation}
	\end{widetext}
	Through the density matrices (\ref{rhofNOON}) and (\ref{rhofNOONpm}), we derive: 
	\begin{equation}
		\left\{
		\begin{array}{l}
			\langle \hat{n} \rangle
			= e^{-\frac{\gamma}{2}T_H}\sin(T_H\delta)
			\\[6pt]
			\frac{\partial \langle \hat{n} \rangle}{\partial \delta}
			= T_H e^{-\frac{\gamma}{2}T_H}\cos(T_H\delta)
			\\[6pt]
			\Delta \hat{n}
			= \sqrt{1 - e^{- \gamma T_H}\sin^{2}(T_H\delta)}
		\end{array}
		\right.
	\end{equation}
	and, calculating again $\Delta \delta = \frac{\Delta \hat{n}}{\left| \partial \langle \hat{n} \rangle / \partial \delta\right|}$, we have now:
	\begin{equation}\label{sensN=1NOON}
		\Delta \delta = \frac{\sqrt{1- e^{- \gamma T_H}\sin^{2}(T_H\delta)}}{\left|  T_H e^{-\frac{\gamma}{2}T_H}\cos(T_H\delta) \right|} .
	\end{equation}
	In Fig.~\ref{FIG2} the result (\ref{sensN=1NOON}) is plotted as a function of the holding time $T_H$ for different values of $\gamma$. Again the noise increasingly degrades the sensitivity as $\gamma$ increases and, for $\gamma \rightarrow 0$, the sensitivity tends to the noise-free function $1/T_H$. The insensitivity points occur here for $T_H=\left( \frac{1}{2}+n\right)\frac{\pi}{\delta}$ ($n \in \mathbb{N}$) and again do not depend on $\gamma$. 
	
	\section{Open interferometer with $N=2$}
	We study here our open quantum interferomenter for $N=2$, again with $\hat{L}=\hat{S}_z$ and $\hat{L}=\hat{S}_{\pm}$ as noise operators.  We performed the analytic calculations of the entire interferometric process computationally. Since they are particularly lengthy and intricate, we do not include them in the discussion but we only show the results. 
	
	\vspace*{0.6cm}
	\subsection{Input state $N0=\ket{2,0}$}
	In the basis $\{ \ket{2,0} ,\ket{1,1}, \ket{0,2} \}$ the initial state $N0=\ket{2,0}$ is described by the matrix
	\begin{equation}\label{rho02}
		\rho_0 = \left(\begin{matrix}
			1&0&0\\0&0&0\\0&0&0
		\end{matrix}\right) .
	\end{equation}
	This input state, in the absence of noise, leads to $\Delta \delta = 1/\sqrt{2}T_H$, which provides our reference behavior for $\Delta \delta$.
	
\begin{figure}[t!] 
		\centering 
		\includegraphics [height=13cm]{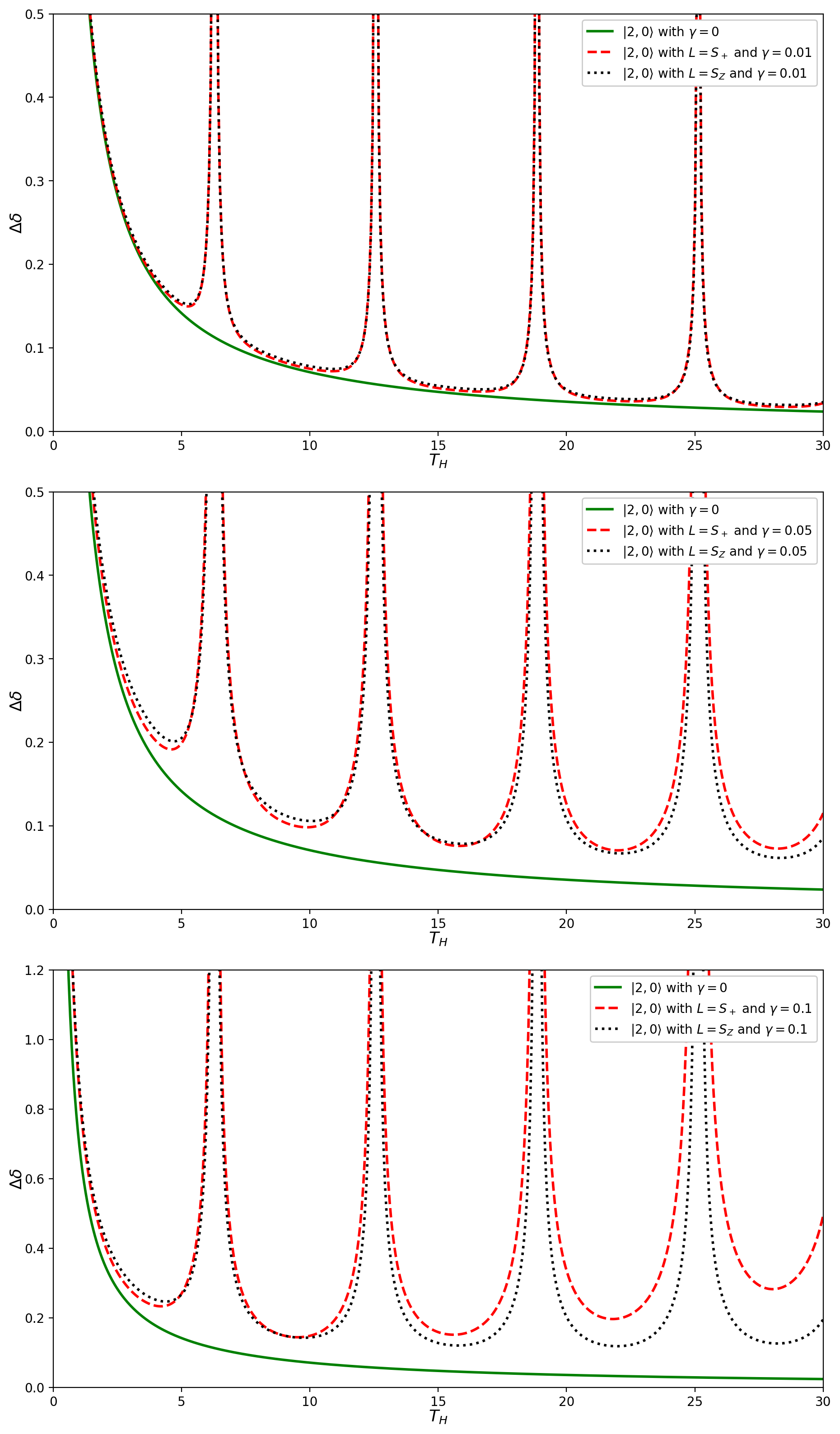} 
		\caption{The sensitivity of the interferometer for input state $\ket{2,0}$ is plotted as a function of the holding time $T_H$ for $\delta=0.5$ and different values of $\gamma$. The cases of noise operators $\hat{S}_z$ and $\hat{S}_{\pm}$ are compared with the noiseless function $1/(\sqrt{2}T_H)$.} 
		\label{FIG3} 
	\end{figure}

	During the first beam splitter the system evolves for a time $T_{BS}=\pi/4$ according to equation (\ref{evut}), where now:
	\begin{equation}
		\hat{H}_J =  \left(\begin{matrix}
			0&-\sqrt{2}&0 \\ -\sqrt{2}&0&-\sqrt{2}\\ 0&-\sqrt{2}&0
		\end{matrix}\right) .
	\end{equation}
	During the phase accumulation stage, the open interferometer undergoes the dissipative evolution (\ref{evlind}), where $\hat{H}_{\delta}$ is here represented by the matrix:
	\begin{equation}
		\hat{H}_{\delta} = \left(\begin{matrix}
			\delta &0&0 \\ 0&0&0 \\ 0&0& - \delta
		\end{matrix}\right) .
	\end{equation}
	\begin{widetext}
		As in the previous section, during the Lindblad evolution we use the noise operators $\hat{S}_z$ and $\hat{S}_{\pm}$, which here read:
		\begin{equation}
			\hat{S}_z = \left(\begin{matrix}
				1&0&0 \\ 0&0&0 \\ 0&0&-1
			\end{matrix}\right), \quad	\hat{S}_{-} = \sqrt{2} \left( \begin{matrix} 0&0&0 \\ 1&0&0 \\ 0&1&0 \end{matrix}\right), \quad \hat{S}_{+} =\sqrt{2} \left(\begin{matrix} 0&1&0 \\ 0&0&1 \\ 0&0&0 \end{matrix}\right) .
		\end{equation}
		At the end of the interferometric process, for $\hat{L}=\hat{S}_z$ we have:
		\begin{equation}
			\Delta \delta (\hat{S}_z) = \frac{\sqrt{3+ e^{-2\gamma T_H} \left(\cos(2T_H \delta) -4e^{\gamma T_H} \cos^2(T_H \delta)\right)}}{\left| -2e^{-\frac{\gamma}{2} T_H} T_H \sin(T_H \delta) \right|} 
		\end{equation} 
		while, for $\hat{L}=\hat{S}_{\pm}$, we obtain
		\begin{equation}\label{30}
			\Delta \delta (\hat{S}_{\pm}) = \frac{\sqrt{e^{-4\gamma T_H }\left(    e^{2\gamma T_H}\left( 1+ \gamma T_H + 2 e^{2\gamma T_H} +  e^{\gamma T_H}\cos(2T_H \delta)\right)        -      \left(1-3e^{\gamma T_H} \right)^2 \cos^2(T_H \delta)         \right)}}{\left| e^{-2\gamma T_H} \left( 3e^{\gamma T_H}-1\right)T_H \sin(T_H \delta)\right| } .
		\end{equation}
	\end{widetext}
	These results are plotted as a function of holding time $T_H$ in Fig.~\ref{FIG3} for different values of $\gamma$. We immediately notice that the sensitivity is actually better in the case $\hat{L}=\hat{S}_z$. The insensitivity points are the same for the three cases $\hat{L}=\hat{S}_z$ and $\hat{L}=\hat{S}_{\pm}$, and they are also the same as those found in Sec.~IV.A when the interferometer had only one particle and input state $\ket{1,0}$. 
	
	\subsection{Input states $TF=\ket{1,1}$ and $NOON$}
	We study here the interferometric process for the initial states $TF=\ket{1,1}$ and $NOON=\frac{1}{\sqrt{2}}\left( \ket{2,0} + \ket{0,2} \right)$. We approach these two cases together because we obtain both leading to the same result. Nevertheless, we emphasize that, \textit{only} for the $NOON$ input state case, we take in the first beam splitter $T_{BS}=\pi/2$.

	At the end of the interferometric process the sensitivity is calculated. The state $\ket{1,1}$ is a perfectly number squeezed state and, when used as input state, it should improve the interferometer sensitivity to the Heisenberg limit. Nevertheless, if no energy shift is present between the wells in the phase accumulation stage, a population imbalance measurement would give no information about phase accumulation. A good estimator is, instead, the parity operator $\hat{\Pi}_b = e^{i\pi \hat{n}_b}$, which in the chosen basis reads:
	\begin{equation}\label{parity}
		\hat{\Pi}_b = \left(\begin{matrix}
			1&0&0 \\ 0&-1&0\\0&0&1
		\end{matrix}\right) . 
	\end{equation}
	Through operator (\ref{parity}) we can calculate $\Delta \hat{\Pi}$ and derive the sensitivity as: $\Delta \delta = \frac{\Delta \hat{\Pi}}{\left| \partial \langle \hat{\Pi} \rangle / \partial \delta\right|}$. 
	The same operator is used in calculating $\Delta \delta$ for the $NOON$ state. As anticipated, for both initial states and considering $\hat{L}=\hat{S}_z$, $\hat{L}=\hat{S}_{\pm}$, we obtain:
	\begin{widetext}
		\begin{equation}\label{33}
			\Delta \delta(\hat{S}_z)= \frac{\sqrt{1- e^{-4\gamma T_H} \cos^2(2T_H\delta)}}{\left|2e^{-2\gamma T_H} T_H\sin(2T_H\delta)\right|} 
		\end{equation}
		and
		\begin{equation}\label{34}
			\Delta \delta(\hat{S}_{\pm}) = \frac{\sqrt{\,1 - e^{-2 \gamma T_H} \left(\cos(2 \delta T_H) - (\gamma T_H)\, e^{-\gamma T_H} \right)^2\,}}{\big|\,2\, e^{- \gamma T_H} \, T_H \, \sin(2 \delta T_H)\,\big|} \,.
		\end{equation}
	\end{widetext}
	
\begin{figure}[t!] 
		\centering 
		\includegraphics [height=13cm]{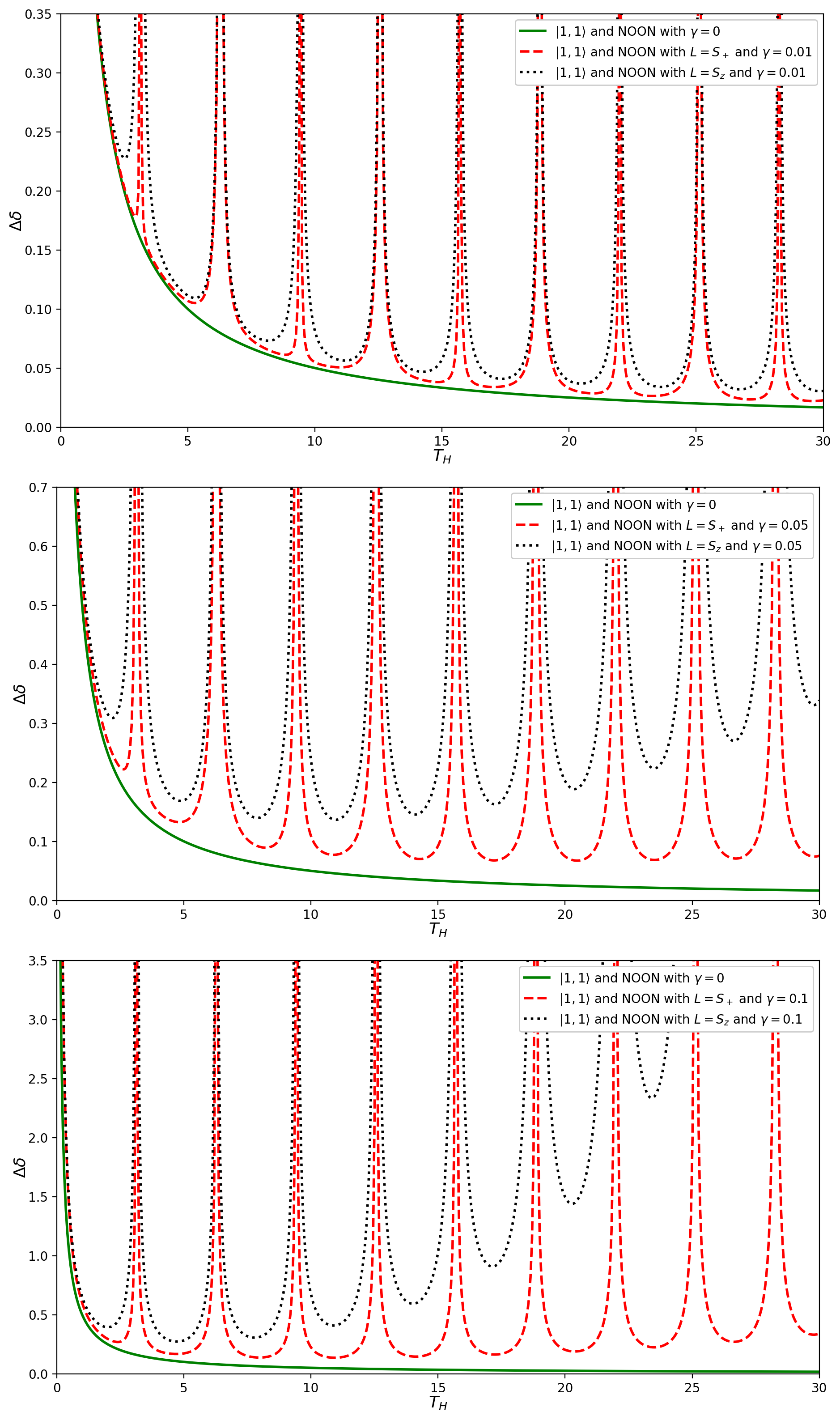} 
		\caption{The sensitivity of the interferometer for input states $\ket{1,1}$ and $NOON$ with $N=2$ is plotted as a function of $T_H$ for $\delta=0.5$ and different values of $\gamma$. The cases of noise operators $\hat{S}_z$ and $\hat{S}_{\pm}$ are compared with the function $1/(2T_H)$.} 
		\label{FIG4} 
	\end{figure}

	The results are given in Fig.~\ref{FIG4}. First of all we notice that, as for the input state $\ket{2,0}$, the insensitivity points do not depend on the value of $\gamma$. In contrast to the case where the input state $\ket{2,0}$ is considered, however, we have here extra insensitivity points provided by the fact that in the denominators of (\ref{33}) and (\ref{34}) there is now $\sin(2T_H \delta)$, leading to singularities at $T_H=\pi n /2 \delta$ ($n \in \mathbb{N}$). The $N=2$ case in this sense is different from $N=1$. In Secs.~IV.A and IV.B we had indeed found the same number of insensitivity points even if shifted by a factor $\pi/2$. In the present case we find instead the points not shifted, but doubled as frequency. Finally, we notice that, in contrast to the $\ket{2,0}$ input state case, the sensitivity is actually less deteriorated for $\hat{L}=\hat{S}_{\pm}$.
	
	\begin{figure*}
		\centering
		\includegraphics[width=\textwidth]{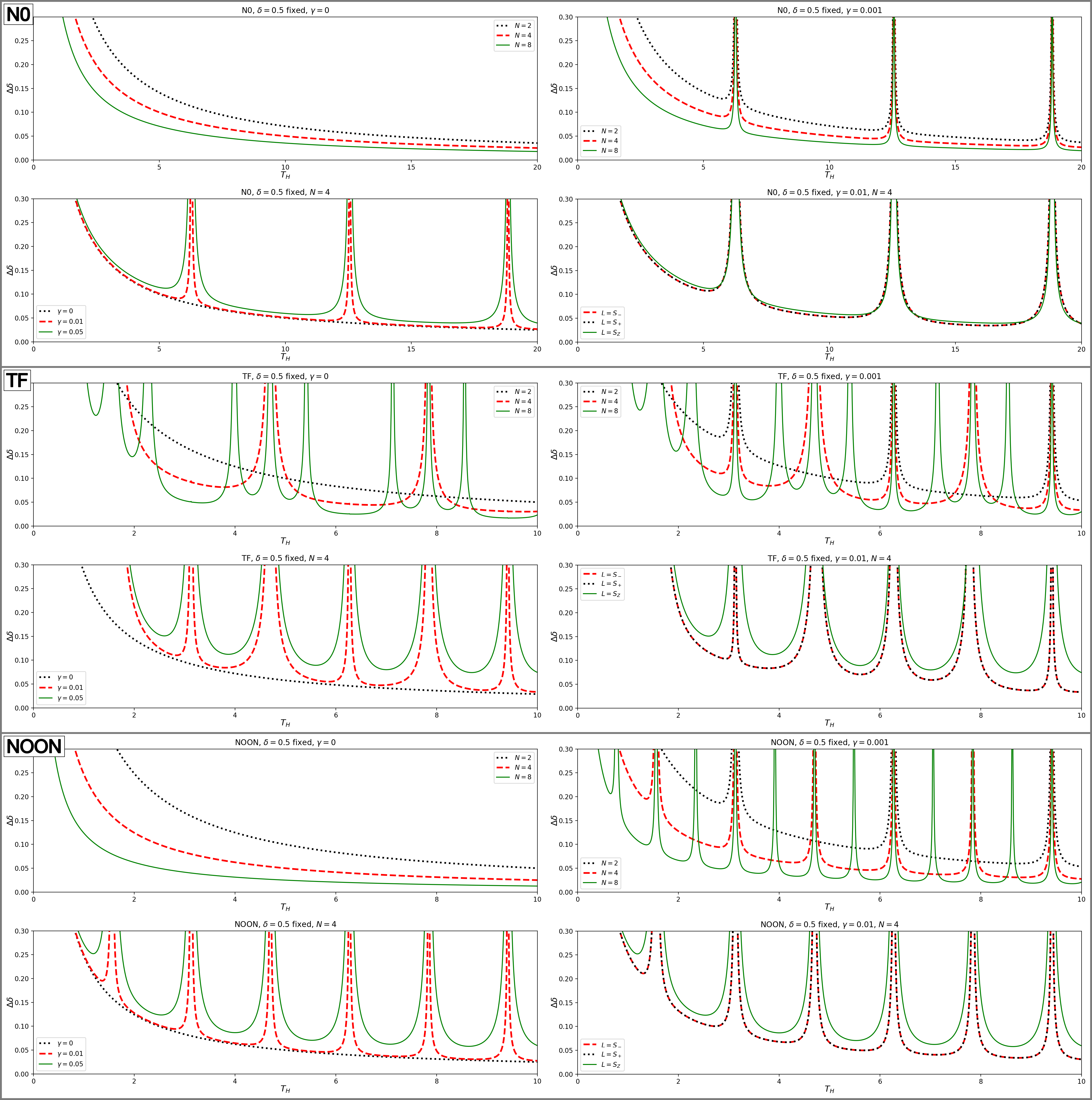}
		\caption{
			The sensitivity $\Delta \delta$ of the atom interferometer for the $N0=\ket{N,0}$ (upper four), $TF=\ket{N/2,N/2}$ (middle four) and $NOON$ (lower four) input states is plotted as function of the holding time $T_H$ for $\delta=0.5$. For each initial state the first plot has fixed $\gamma=0$ and $\Delta \delta$ is plotted at different $N$. The second plot shows $\gamma \neq 0$ plots at different $N$. In the third graph $N=4$ is fixed and different plots differ in their $\gamma \neq 0$ parameter. In the previous three plots we always fixed the Lindblad operator to $\hat{S}_z$. The fourth plot has $N=4$ and $\gamma=0.01$, both fixed, and $\Delta \delta$ is plotted for different Lindblad operators $\hat{S}_-$, $\hat{S}_+$ and $\hat{S}_z$. We see the insensitivity points mostly appear only for the noisy cases (with the exception of the $\ket{N/2,N/2}$ case). Increasing $\gamma$ never results in shifts of the insensitivity points positions in $T_H$. Comparing different Lindblad operators we see that the $\hat{S}_-$ and $\hat{S}_+$ noise operators yield the best sensitivity behavior for every initial state. 
		}
		\label{fig:8}
	\end{figure*}

	\section{Open interferometer with $N>2$ }
	
	Having obtained the results for the emergence and structure of the insensitivity points in the $N=1$ and $N=2$ cases, we now turn our attention to how these features extend to larger particle numbers $N$. In order to provide a more complete and systematic comparison of the usefulness of different initial states in atom interferometry, we focus on several aspects of the sensitivity behavior. First, we examine whether insensitivity points appear exclusively in the presence of noise, or whether they also occur in noiseless scenarios when $N>2$. Second, since for $N=1,2$ the positions of the insensitivity points remain essentially unaffected by changes in the noise intensity, we investigate whether this robustness persists for larger $N$. Third, we analyse how the number--density of insensitivity points in the holding time $T_H$ changes with increasing $N$. Finally, we study how the sensitivity in the most favorable regions---namely, values of $T_H$ sufficiently far from any insensitivity point---scales with $N$ for both number-conserving and particle non-conserving Lindblad operators.
	
	For $N>2$, where obtaining the exact analytical solution of the Lindblad master equation becomes computationally prohibitive, we rely entirely on numerical methods to evaluate the system's evolution. Our strategy is as follows. We first compute the unitary operator corresponding to the beam--splitter by numerically exponentiating the beam--splitter Hamiltonian (\ref{HJ}), and apply it to the chosen initial state. We then evolve the resulting density matrix during the holding stage---where the dynamics is non-unitary---by employing a fourth-order Runge--Kutta algorithm, repeating the procedure for different values of the holding time $T_H$. After the holding stage, we apply once more the beam--splitter unitary transformation, thereby obtaining the final density matrix. From this final state we evaluate the sensitivity using the standard expressions:
	\begin{equation}
		\Delta \delta = \frac{\Delta \hat{n}}{\left| \partial \langle \hat{n} \rangle / \partial \delta \right| }\quad \text{or} \quad \frac{\Delta \hat{\Pi}_b}{\left| \partial \langle \hat{\Pi}_b \rangle / \partial \delta \right| }\, .
	\end{equation}

	In the following subsections, we analyze each of the aspects listed above separately, considering the same three initial states as in the $N=2$ case. We begin with the $N0=\ket{N,0}$ state, then we examine the $TF=\ket{\frac{N}{2},\frac{N}{2}}$ state, and finally we study the $NOON$ state. For each initial state, we present four sets of graphs. The first two show how the sensitivity depends on the holding time $T_H$ for different values of $N$, in the absence and presence of noise, respectively. The third graph focuses on the $N=4$ case for different noise intensities, while the fourth compares different choices of Lindblad operators (again for $N=4$) to model the noise. Lastly, we study the introduction of a particle non-conserving Lindblad operator and the $N$--scaling of the sensitivity. 

	\subsection{Initial state $N0=\ket{N,0}$}
	We start by analysing the initial state $\ket{N,0}$. In Fig.~\ref{fig:8} we see that there are no insensitivity points for the noise-free case, for any $N$. When the noise parameter $\gamma \neq 0$, the insensitivity points emerge at holding times $T_H = 2\pi n$, where $n$ is a positive integer. This remains true for any number of particles $N$, as well as for any $\gamma>0$ we considered. As expected, increasing $\gamma$ worsens the sensitivity. Lastly, the positions of insensitivity points are also independent of which Lindblad operator is used among those considered.

	\subsection{Initial state $TF=\ket{\frac{N}{2},\frac{N}{2}}$}
	Next, we analyse the initial state $\ket{\frac{N}{2},\frac{N}{2}}$. In Fig.~\ref{fig:8} we see that there are insensitivity points already in the noise-free case, for $N\geq4$. Moreover, their number increases with increasing $N$ as $(N-2)/2$ per period $\pi$ in $T_H$. When the noise parameter $\gamma$ is turned on, additional insensitivity points emerge at holding times that are integer multiples of $\pi$. Further increasing $\gamma$ or choosing a different Lindblad noise operator does not alter their position; however, the latter can improve the sensitivity. Again, increasing $\gamma$ worsens the sensitivity.

	\subsection{Initial state $NOON$}
	
	Finally, we analyse the initial state $NOON$. Much like in the $\ket{N,0}$ case, there are no insensitivity points in the noise-free case. However, when the noise parameter $\gamma$ is turned on, they emerge at positions $T_H = 2 \pi / N$. Again, further increasing $\gamma$ or choosing a different Lindblad noise operator does not alter their position, and increasing $\gamma$ worsens the sensitivity. We also note that the noise operators $\hat{S}_-$ and $\hat{S}_+$ yield better sensitivity than $\hat{S}_z$.

	\subsection{Particle non-conserving Lindblad operator}
	So far we have constrained ourselves to number-conserving Lindblad operators: $\hat{S}_{-}$, $\hat{S}_{+}$ and $\hat{S}_{z}$. We now consider the behavior of the sensitivity when $\hat{L}=\frac{1}{\sqrt{2}}(\hat{a} +\hat{b}) \equiv \hat{\alpha}$. In this case the evolution of the quantum state is not constrained to states with a definite number of particles $N$ of the initial state, but instead results in population of states with lower $N$ as well. Therefore, it is necessary to simulate the whole Fock space up-to the initial state particle number $N$. 
	
\begin{figure} 
		\centering 
		\includegraphics[width=0.45\textwidth]{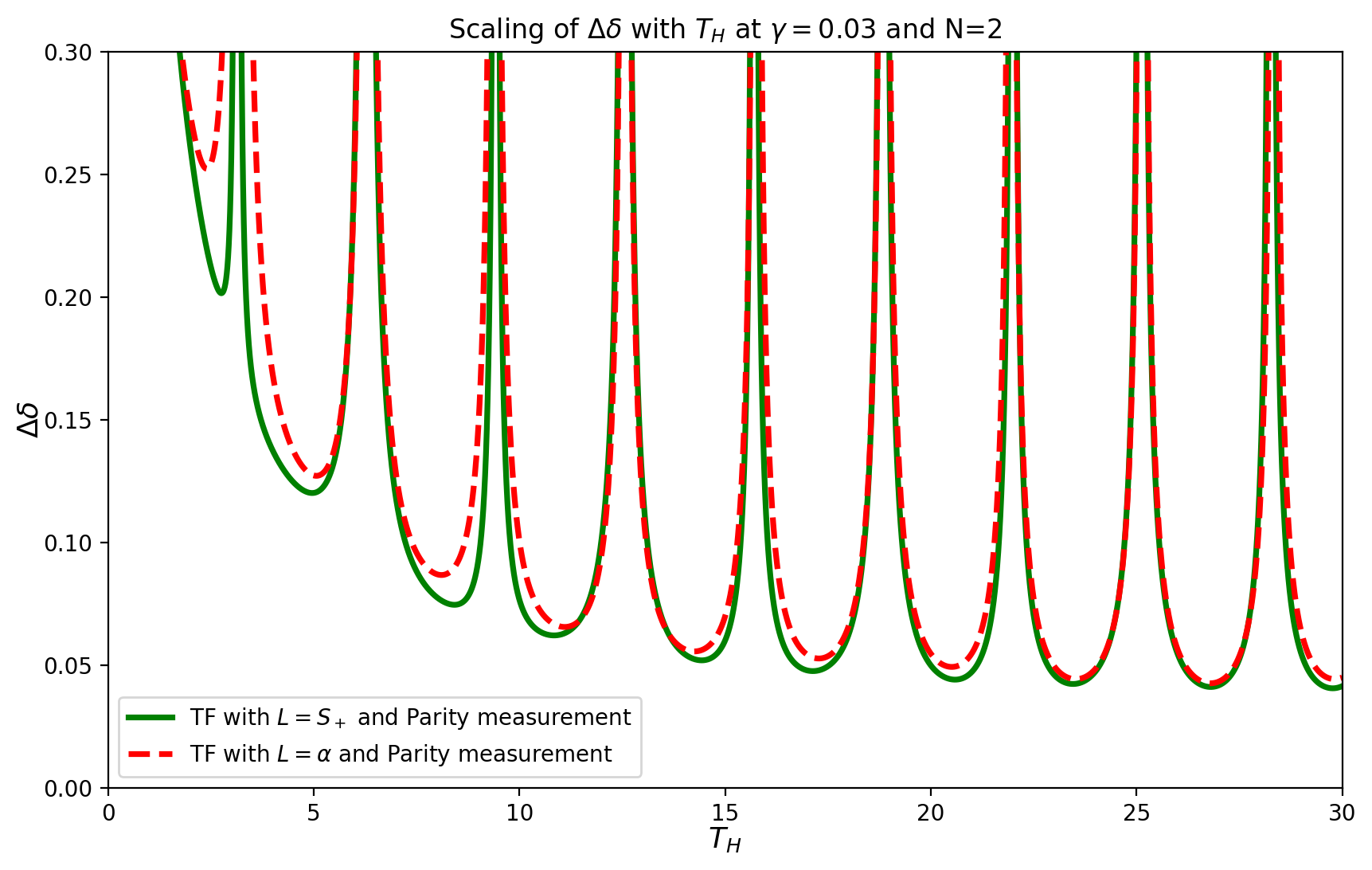}
		\caption{
			The sensitivity $\Delta \delta$ for the $TF=\ket{N/2,N/2}$ input state is plotted as a function of the holding time $T_H$ for $\delta=0.5$ at $N=2$ and $\gamma=0.03$. The dashed and solid lines represent the particle non-conserving $\hat{L}=\hat{\alpha}$ and the particle conserving $\hat{L}=\hat{S}_+$, respectively. The measurement operator corresponding to TF state is the parity operator $\hat{\Pi}_b$.
		} 
		\label{fig:10} 
	\end{figure}

	First we compare the behavior of the sensitivity $\Delta \delta$ with the holding time $T_H$ between the evolutions caused by $\hat{L}=\hat{S}_+$ and $\hat{L}=\hat{\alpha}$ for TF inital state and $N=2$ in Fig.~\ref{fig:10}. We see that for the chosen value of the noise parameter $\gamma = 0.03$, evolution with $\hat{L}=\hat{S}_+$ results in a better sensitivity on the whole range of $T_H$ considered.
	
\begin{figure} 
		\centering 
		\includegraphics[width=0.45\textwidth]{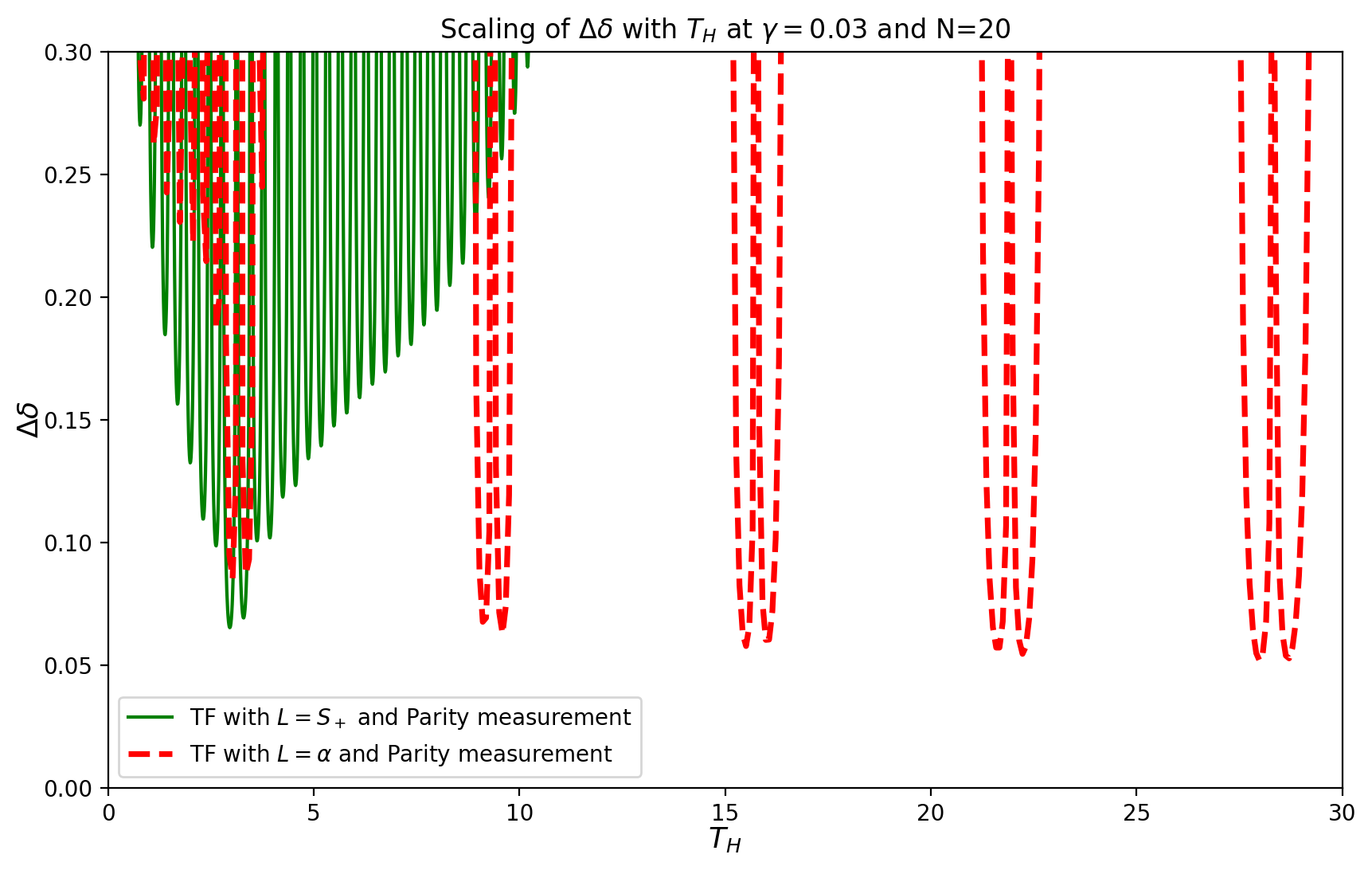}
		\caption{
			The sensitivity $\Delta \delta$ for the $TF=\ket{N/2,N/2}$ input state is plotted as a function of the holding time $T_H$ for $\delta=0.5$ at $N=16$ and $\gamma=0.03$. The dashed and solid lines represent the particle non-conserving $\hat{L}=\hat{\alpha}$ and the particle conserving $\hat{L}=\hat{S}_+$, respectively. The measurement operator corresponding to TF state is the parity operator $\hat{\Pi}_b$.
		} 
		\label{fig:11} 
	\end{figure}

	Increasing $N$ to $20$ for the initial state, reveals that the sensitivity starts to deteriorate for the number-conserving $\hat{L}=\hat{S}_+$ with $T_H$ (Fig.~\ref{fig:11}). On the other hand, the sensitivity for $\hat{L}=\hat{\alpha}$ keeps steadily improving with $T_H$. Nevertheless, in the considered range of $T_H$ the minimum achieved with $\hat{L}=\hat{S}_+$ outperforms the results obtained with $\hat{L}=\hat{\alpha}$ for all other values of $T_H$.
	
	\subsection{Scaling of the sensitivity with $N$}
\begin{figure} 
		\centering 
		\includegraphics[width=0.45\textwidth]{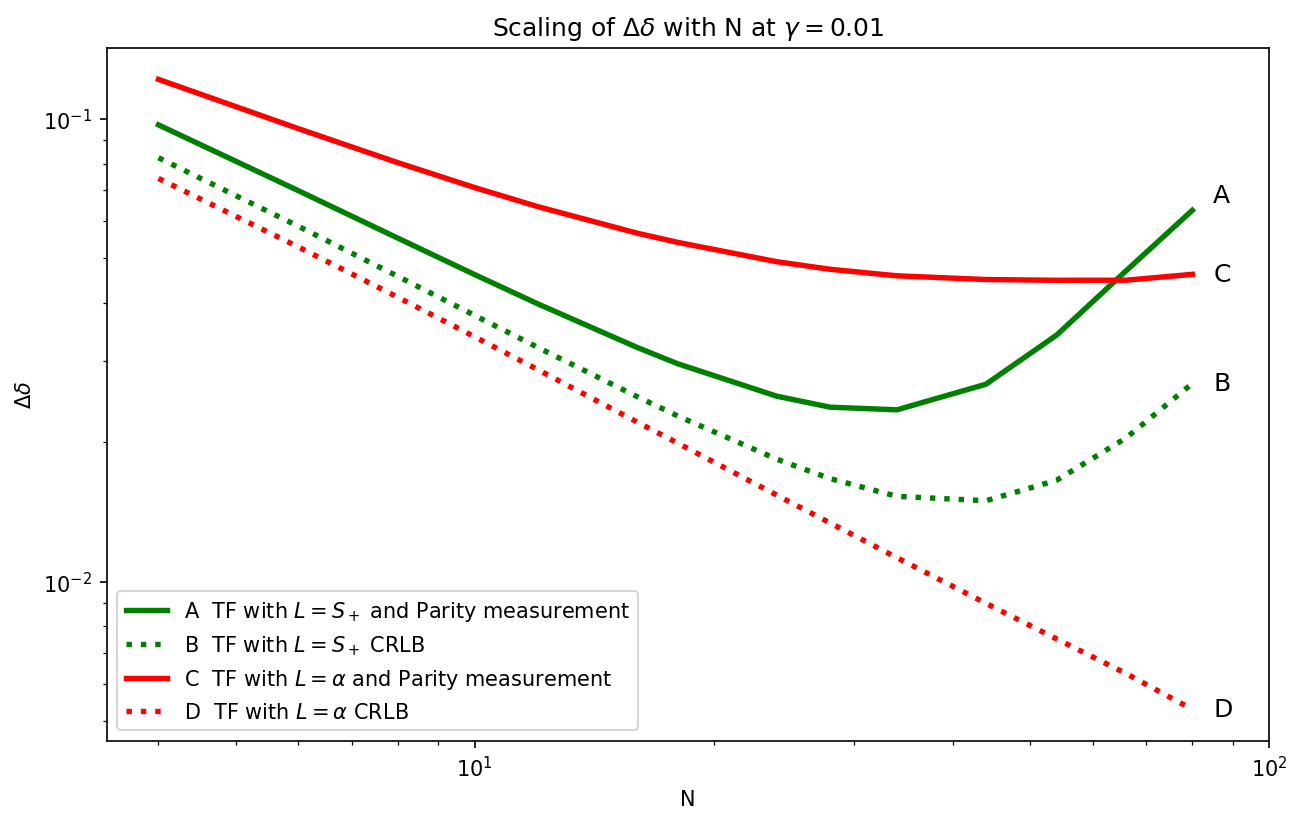}
		\caption{
			The sensitivity $\Delta \delta$ for the $TF=\ket{N/2,N/2}$ input state is plotted as a function of the initial particle number $N$ for $\delta=0.5$ and $\gamma=0.03$ at the holding time corresponding to the first minimum after $T_H = \pi$ (solid lines A and C). Dotted lines B and D represent the CRLB to sensitivity $\Delta \delta$ with respect to the end measurement operator. The pairs of lines A-B and C-D encode the particle non-conserving operator $\hat{L}=\hat{\alpha}$ and the particle conserving $\hat{L}=\hat{S}_+$, respectively.  
		} 
		\label{fig:12} 
	\end{figure}

	The difference in the nature of the sensitivity behavior with respect to the holding time between the particle conserving and particle non-conserving Lindblad operators prompts us to investigate further the scaling of the sensitivity minima with increasing $N$. For this we focus again on the TF initial state. In Fig.~\ref{fig:12} we plot the scaling of the minima of the sensitivity (i.e. points in $T_H$ far from the insensitivity points) with $N$. As those minima can appear at different holding times $T_H$, we will focus on the first minimum that appears at or after $T_H = \pi$ as a rule of the protocol. Furthermore, we always choose the parity operator as the end measurement used to estimate the sensitivity. The red and green solid lines represent the dependence of sensitivity on the number of particles $N$ for Liouvillian evolutions generated by $\hat{L}=\hat{\alpha}$ and $\hat{L}=\hat{S}_+$ respectively. We see that at a certain $N$ the sensitivity corresponding to $\hat{L}=\hat{\alpha}$ evolution starts to outperform the one corresponding to $\hat{L}=\hat{S}_+$. As this measurement protocol and its results depend strongly on the specific end–measurement employed, we also include, as dotted lines, the corresponding CRLB for both noise models and for each value of $N$. The CRLB represents the ultimate sensitivity attainable with an optimal measurement. From the dotted red curve we indeed see that, under evolution generated by $\hat{L}=\hat{\alpha}$, the optimal sensitivity would surpass that achieved with $\hat{L}=\hat{S}_+$ for all $N$, were one to adopt the measurement that saturates the bound at each point.

	\section{Conclusions}
	
	In this paper we have characterized the emergence and behavior of the insensitivity points arising in a two–mode interferometer when it is coupled to an external environment. Throughout the main text we assumed that the environment couples to the system only during the phase–accumulation stage, and we focused on three representative input states: $N0 \equiv \ket{N,0}$, $TF \equiv \ket{N/2,N/2}$ and $NOON \equiv \frac{1}{\sqrt{2}}\left(\ket{N,0}+\ket{0,N}\right)$ in the presence of number-conserving and non-conserving Lindblad operators.
	
	For $N=1$ and input $\ket{1,0}$ we found that, for the considered number-conserving noise operators $\hat{L}=\hat{S}_z$ and $\hat{L}=\hat{S}_{\pm}$, the sensitivity as a function of $T_H$ varies as:
	\begin{equation}\label{fin}
		\Delta \delta =
		\frac{\sqrt{1 - e^{- \gamma T_H}\cos^{2}(T_H\delta)}}
		{\left| - T_H e^{-\frac{\gamma}{2}T_H}\sin(T_H\delta) \right|} .
	\end{equation}
	This expression illustrates a key feature of the open interferometer: while the noise strength $\gamma$ affects the overall sensitivity, the locations of the divergence points---the insensitivity points---remain independent of~$\gamma$. We found them to depend on the input state, as expected.
	
	For $N=2$ we showed that the sensitivity turns out to be different for the cases $\hat{L}=\hat{S}_z$ and $\hat{L}=\hat{S}_{\pm}$. Again, we found that the insensitivity points do not depend on $\gamma$, but still depend on the input state.
	
	Using numerical techniques we obtained the behavior of $\Delta \delta$ for $N>2$. Unlike the $N=1$ and $N=2$ cases, we observed that the insensitivity points can emerge in the noise-free regime as well (e.g.\ the $TF$ state with $N=4$ particles). Next, we confirmed that the position of the insensitivity points does not change with the noise intensity $\gamma$ when considering number-conserving Lindblad operators. More interestingly, we found that the density of the insensitivity points grows linearly with $N$ for the $TF$ and $NOON$ states. On the other hand, for the $N0$ state, the number of insensitivity points remains the same for each $N$. This is a consequence of the $N0$ state being separable, and therefore equivalent to $N$ copies of $\ket{1,0}$, i.e.\ $\ket{N,0} = \bigotimes_{i=1}^N \ket{1,0}_i$.
	
	We see that, for our particular choice of the measurement operator $\hat{\Pi}_b$ and the $TF$ initial state, the sensitivity obtained with the number-conserving noise operator performs better in terms of sensitivity than that obtained with the particle non-conserving operator for small $N$, while the opposite occurs for larger values of $N$. However, both sensitivities deteriorate and start to worsen with $N$ in the limit of large particle numbers. We emphasise that this behavior is a consequence of our specific choice of the measurement operator used at the end of the protocol. If instead we consider the Cram\'{e}r–Rao lower bound, which gives the best obtainable sensitivity, we see that the particle non-conserving case always performs better than the number-conserving one. Furthermore, the latter displays a minimum at certain system-specific values of $N$, after which it starts to deteriorate. By contrast, the particle non-conserving case keeps improving even at large $N$, albeit only at a shot-noise rate, in agreement with the results of \cite{Davidovich2011}.


	\section*{Acknowledgements}
	We acknowledge Fabio Benatti and Luca Pezz\`e for helpful and useful discussions.
	T.F. and A.T. thank the Project \lq\lq National Quantum Science and Technology Institute – NQSTI\rq\rq \, Spoke 3: \lq\lq\ Atomic, molecular platform for quantum technologies\rq\rq. \v{Z}.K. and A.T. acknowledge funding from the European Union’s Horizon Europe Research and Innovation Programme under the Marie Sk\l{}odowska-Curie Doctoral Network MAWI under the grant agreement No. 101073088.


	\appendix
	
	\begin{figure}[t!] 
		\centering 
		\includegraphics [height=16.5cm]{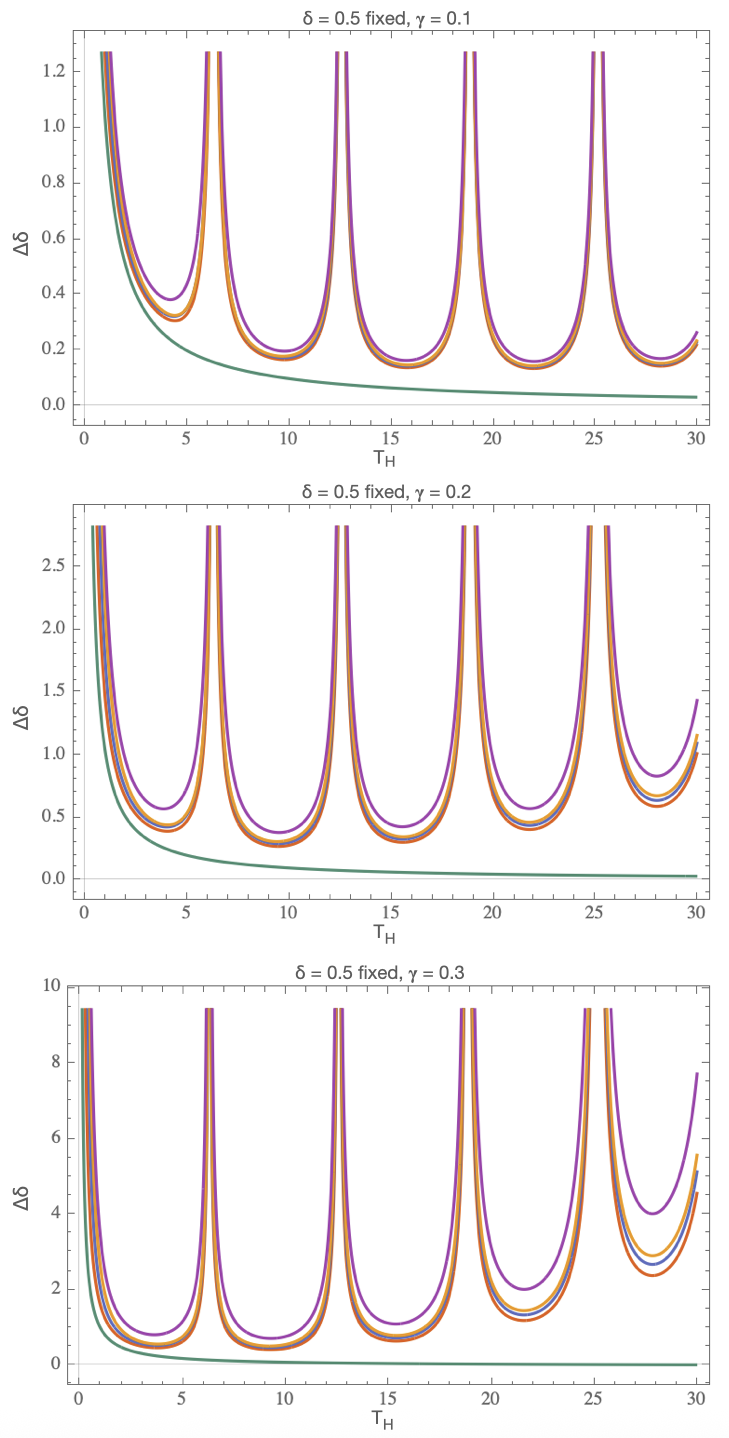} 
		\caption{The sensitivity of the interferometer for input states $\ket{1,0}$ and noise present throughout the whole process is plotted as a function of $T_H$ with $\delta=0.5$. In each panel, the curves are listed in the order in which they appear from bottom to top (corresponding to increasing degradation of the sensitivity): the noise-free function $1/T_H$, the result obtained in the main text where the noise is present only during the phase accumulation stage, then the cases of $\hat{S}_z$, $\hat{S}_{-}$ and $\hat{S}_{+}$.}
	\label{lastFIG} 
\end{figure}

\section{Noise present throughout the full process}
We show here the behavior of the sensitivity when the dynamic of the interferometer is affected by the noise operators $\hat{L}=\hat{S}_{z}$ and $\hat{L}=\hat{S}_{\pm}$ throughout the full interferometric process and not only during the phase accumulation stage. We discuss the case with $N=1$ and input state $\ket{1,0}$, because analytical calculations, although done computationally, become too lengthy for $N=2$. Nevertheless, also for $N=1$, we will not show the equations, but we will provide the results graphically.

Being all the process affected by the noise, during the first beam splitter the dynamics of the interferometer is governed for a time $T_{BS}=\pi/4$ ($J=1$) by the equation
\begin{equation}\label{evlindJ}
	\frac{\partial \rho}{\partial t} = - i\left[\hat{H}_{J},\rho\right] + \gamma \hat{L} \rho \hat{L}^{\dagger} - \frac{\gamma}{2}\left\{\hat{L}^{\dagger} \hat{L} ,\rho\right\}
	\\ [1.5em]
\end{equation}
where:
\begin{equation}
	\hat{H}_J = \left( \begin{matrix}
		0 & -1 \\ -1 & 0
	\end{matrix}\right) .
\end{equation}
During the phase accumulation stage the system undergoes the dynamics already described in the main text, namely
\begin{equation}
	\frac{\partial \rho}{\partial t} = - i\left[\hat{H}_{\delta},\rho\right] + \gamma \hat{L} \rho \hat{L}^{\dagger} - \frac{\gamma}{2}\left\{\hat{L}^{\dagger} \hat{L} ,\rho\right\}
\end{equation}
where the Hamiltonian reads:
\begin{equation}
	\hat{H}_{\delta} = \left(\begin{matrix}
		\delta/2&0\\0&-\delta/2
	\end{matrix}\right) . 
\end{equation} 
The second beam splitter is obtained restoring the symmetry between the wells and letting the particle to tunnel again following (\ref{evlindJ}) for a time $T_{BS}=\pi/4$. At the end of the interferometric process the sensitivity is derived through the operator $\hat{n}$, namely: $\Delta \delta = \frac{\Delta \hat{n}}{\left| \partial \langle \hat{n} \rangle / \partial \delta\right|}$. 

The results are given in Fig.~\ref{lastFIG}. We immediately notice that, unlike the case of noise present only in the phase accumulation stage, the sensitivity obtained considering the three operators $\hat{L}=\hat{S}_{z}$, $\hat{L}=\hat{S}_{\pm}$ is no longer equal, but differs slightly. The operator $\hat{L}=\hat{S}_{-}$ turns out to be the one that worsens sensitivity the most. In the figure the present case is also compared with the one of Sec.~IV, where the sensitivity for all the noise operators was given by equation (\ref{sensN=1}) (or equivalently (\ref{fin})). It can be seen, as expected, that assuming the noise throughout the entire process leads to a slight worsening of the sensitivity. The effect is smaller for $\hat{L}=\hat{S}_{z}$ and $\hat{L}=\hat{S}_{+}$, while it is more pronounced for $\hat{L}=\hat{S}_{-}$.These differences grow as $\gamma$ and the holding time $T_H$ increase. Finally, we notice and emphasize that the insensitivity points do not change. They are the same for the three operators and they are also the same as those encountered in Sec.~IV.


\bibliographystyle{apsrev4-1}   
\bibliography{references}       

\end{document}